# Artificial Intelligence for Pediatric Height Prediction Using Large-Scale Longitudinal Body Composition Data

[1,2]Dohyun Chun[*], [3]Hae Woon Jung[*], [4]Jongho Kang, [5,6]Woo Young Jang[†], [2,7,8]Jihun Kim[‡]

[1]*College of Business Administration, Kangwon National University, Chuncheon, Gangwon-do, Korea.*

[2]*Research Team, The Global Prediction Co., Ltd., Gyeonggi-do, Gwangmyeong-si, Korea*

[3]*Department of Pediatrics, Kyung Hee University Medical Center, Seoul, Korea*

[4]*College of Business Administration, Chonnam National University, Gwangju, Korea*

[5]*Institute of Nano, Regeneration, Reconstruction, Korea University, Seoul, Korea*

[6]*Department of Orthopedic Surgery, College of Medicine, Korea University, Seoul, Korea*

[7]*College of Humanities & Social Sciences Convergence, Yonsei University, Wonju, Gangwon-do, Korea.*

[8]*Yonsei Institute of AI Data Convergence Science, Yonsei University, Wonju, Gangwon-do, Korea.*

**Abstract**

**Objective**: This study aimed to develop an accurate and reliable artificial intelligence (AI) model for predicting future height in children and adolescents using anthropometric and body composition.

**Materials and Methods**: We utilized a comprehensive longitudinal dataset from the GP Cohort Study, encompassing 588,546 measurements from 96,485 children and adolescents aged 7-18. The prediction model incorporated anthropometric and body composition measures, their standard deviation scores (SDS), and velocity parameters for height and body composition. To evaluate model performance, we employed root mean square error (RMSE), mean absolute error (MAE), and mean absolute percentage error (MAPE). The model's interpretability was enhanced through various explainable AI techniques, including feature importance analysis, accumulated local effects plots, and Shapley additive explanations (SHAP).

**Results**: The model demonstrated high accuracy in predicting future height. For males, the average RMSE, MAE, and MAPE were 2.51 cm, 1.74 cm, and 1.14%, respectively. Female predictions showed similar accuracy, with RMSE, MAE, and MAPE of 2.28 cm, 1.68 cm, and 1.13%, respectively. Explainable AI approaches demonstrated that SDS of height, height velocity, and soft lean mass velocity were crucial predictors of future height. The model generated personalized growth curves by estimating individual-specific height trajectories, comparing these projections with actual measurements, and displaying key variables through local SHAP.

---

[*] These authors contributed equally: Dohyun Chun, Hae Woon Jung
[†] Corresponding author. E-mail address: opmanse@gmail.com
[‡] Corresponding author. E-mail address: jihunkim79@gmail.com

**Conclusion**: We developed an innovative machine learning model for predicting future height in children and adolescents, utilizing comprehensive anthropometric and body composition data from a large-scale longitudinal study. Our model demonstrates high accuracy in height prediction and generates personalized growth curves, while incorporating explainable AI techniques to enhance interpretability. This approach advances pediatric growth assessment, offering a robust tool for clinical decision support. Despite certain limitations, our method shows promise in early identification of growth disorders and optimization of growth outcomes.

---

## 1. Introduction

Growth assessment and height prediction in children and adolescents are crucial components of pediatric care. Height growth serves as a key health indicator, reflecting the interplay of genetic, environmental, and socioeconomic factors (Norris et al., 2022; Baxter-Jones et al., 2011; Hargreaves et al., 2022). Monitoring height growth enables early detection of disorders, facilitating timely interventions (Saari et al., 2015; Craig et al., 2011; Grote et al., 2008; Zhang et al., 2016). Accurate future height prediction is essential for diagnosing growth disorders, initiating hormone therapy, and evaluating treatment efficacy (Collett-Solberg et al., 2019; Ostojic, 2013; Cuttler & Silvers, 2004). Traditional height prediction methods rely on skeletal maturity assessment using hand-wrist radiographs. These include the Bayley-Pinneau (Bayley and Pinneau, 1952), Tanner-Whitehouse (Tanner et al., 1975), and Roche-Wainer-Thissen (Roche et al., 1975) methods. However, these approaches have limitations including radiation exposure, the need for specialized expertise, and high interobserver variability (Bull et al., 1999; Chávez-Vázquez et al., 2024; Prokop-Piotrkowska et al., 2021). Moreover, their accuracy decreases in children with growth disorders or those undergoing growth hormone therapy (Topor et al., 2010; Bramswig et al., 1990; Bonfig & Schwarz, 2012).

Given the limitations of traditional growth assessment methods, body composition measures have emerged as crucial complementary indicators for monitoring and predicting pediatric growth and development (Johnson et al., 2012; Dalskov et al., 2013; Jamison et al., 2006; Alimujiang et al., 2018; He and Karlberg, 2001; Brener et al., 2017). Recent advances in body composition assessment techniques, such as multifrequency bioelectrical impedance analysis (BIA), have enabled large-scale and long-term pediatric data collection (Chun et al., 2024a; Jansen et al., 2018; Hanem et al., 2019; Rodríguez-Carmona et al., 2022; Atazadegan et al., 2023). Machine learning (ML) techniques are suitable for analyzing these large-scale longitudinal datasets in growth modeling and height prediction. In growth-related fields, ML has been primarily applied to bone age assessment through image processing (Chávez-Vázquez et al., 2024; Martin et al., 2016; Thodberg et al., 2009, 2012; Suh et al., 2023; Satoh et al., 2015). Some studies have attempted direct height prediction using ML, but these relied solely on basic anthropometric data without incorporating body composition measures (Mlakar et al., 2023; Rativa et al., 2018; Shmoish et al., 2021).

This study addresses the current gap in pediatric growth prediction by leveraging ML algorithms on a large-scale longitudinal dataset of integrated anthropometric and body composition measures. Our approach employs big data processing techniques to capture complex relationships between growth-related variables and height growth. We aim to develop accurate ML models for height prediction, identify key predictors, and generate personalized growth curves. Our method offers several unique advantages. First, we incorporate SDS for body composition measures, derived from a comprehensive body composition database (Chun et al., 2024b). These SDS quantify a relative growth status within age- and sex-matched cohorts, offering valuable insights beyond raw measurements. Second, our longitudinal data enables the calculation of growth velocities, a crucial factor often absent in cross-sectional studies. Growth velocities in height and body composition are essential for identifying deviations from normal growth patterns (WHO 2009; Voss et al., 1991; McCarthy et al., 2007; Ekelund et al., 2006). The inclusion of body composition velocity offers unique insights into children's dynamic physical development, enabling a more comprehensive assessment of growth trajectories. Finally, we enhance model interpretability using explainable AI techniques, including Shapley additive explanations (SHAP), to analyze critical factors driving growth predictions (Karim et al., 2023; Gu et al., 2020; Liu et al., 2023; Javidi et al., 2024; Moncada-Torres et al., 2021; Seethi et al., 2024). By integrating personalized growth curve estimation with local interpretability, our model facilitates early identification of at-risk children and enables targeted interventions, complementing traditional growth monitoring methods.

The remainder of this paper is organized as follows: Section 2 describes the data and methods. Section 3 presents the results of our height prediction model and explainable AI analyses. Section 4 discusses our findings and their implications. Finally, Section 5 concludes the paper and suggests future research directions.

## 2. Data and Methods

### 2.1. Study Population

The AI model was trained on data from the GP Cohort Study, a longitudinal investigation conducted by Global Prediction Co., Ltd. (GP), a growth research company in Gwangmyeong, South Korea. This study includes students from elementary, middle, and high schools (aged 7–18 years) in Gyeonggi Province, with an average of 35 schools participating each year. The study commenced on January 1, 2013; since that time, the study has collected 588,546 data points from 96,485 children and adolescents (50,480 males and 46,005 females) born from 1998–2016. All the included children and their guardians provided informed consent, and those unable to tolerate standard measurement methods or who filed to provide informed consent were excluded. The Institutional Review Board (IRB) of Korea University Anam Hospital approved the study. Due to its retrospective nature and the use of de-identified data, the IRB waived the requirement for informed consent (2025AN0110).

*2.2. Data Collection and Measurements*

Trained examiners visit the schools twice yearly to obtain measures of all participating students' heights, weights, and body composition using a standardized protocol. Extensively trained researchers used an octopolar multifrequency BIA (Inbody models J10 and J30, Inbody Inc., Seoul, Korea) and followed established standard operating procedures. Height and body composition were measured using a stadiometer and InBody Body Composition Analyzers. The researchers who collected cohort data were not involved in the subsequent statistical analysis.

Predictor variables of interest included height, weight, protein mass, soft lean mass (SLM), body fat mass (BFM), skeletal muscle mass (SMM), bone mineral mass, total body water (TBW), basal metabolic rate (BMR), and waist and hip circumferences. Body mass index (BMI), body fat mass index (BFMI), skeletal muscle mass index (SMMI), and waist-to-hip ratio (WHR) were calculated as follows:

$$BMI\ (kg/m^2) = \frac{Weight\ (kg)}{Height\ (m)^2},$$

$$BFMI(kg/m^2) = \frac{BFM\ (kg)}{Height\ (m)^2},$$

$$SMMI(kg/m^2) = \frac{SMM\ (kg)}{Height\ (m)^2},$$

$$WHR = \frac{Waist\ circumference\ (cm)}{Hip\ circumference\ (cm)},$$

Anthropometric measurements of children and adolescents were standardized using the 2017 Korean National Growth Charts. This included height, weight, and body mass index. Body composition parameters not available in these charts were standardized using the GP growth chart developed by Chun et al. (2024b). The dataset incorporated both original measurements and their corresponding age- and sex-specific standard deviation scores (SDS).

*2.3. Data Preparation*

After data preprocessing, we included 276,301 measurements (145,292 for males and 133,009 for females) from 54,374 children and adolescents (28,592 males and 25,782 females) aged 7-16. Each measurement encompassed 27 attributes, including basic variables like sex and age, anthropometric and body composition measures, and associated SDS.[1]

The prediction target was defined as the future height growth rate, derived as follows. The prediction dataset construction involved identifying entities with multiple observations. For each entity, two observations were selected: observation 1 and observation 2, with the latter occurring later in time. The height from observation 2 (Height_T) became the prediction target based on observation 1 data. We denoted the age in months for observation 2 as Months_T. The growth rate, calculated as

$$Growth\_Rate = \frac{Height\_T - Height}{Height} \times 100 (\%),$$

served as the target variable for observation 1. Months_T was incorporated into the predictor variables for observation 1. Figure 1 illustrates an example of the resulting prediction dataset for a specific entity, emphasizing the inclusion of target variables. This process, applied to entities with two or more observations, resulted in a dataset of 44,683 entities (23,402 males and 21,281 females) encompassing 206,112 observations (106,461 for males and 99,651 for females).

**Figure 1.** Example of a Prediction Dataset with Target Variables

| ID | Sex | Months | Height | Weight | ... |
|---|---|---|---|---|---|
| 1 | Female | 88 | 122.1 | 21.5 | |
| 1 | Female | 93 | 124.2 | 23.1 | |
| 1 | Female | 136 | 142.7 | 32.9 | |
| 1 | Female | 142 | 146.4 | 34.1 | |

→

| ID | Sex | Months | Height | Weight | ... | Height_T | Growth_Rate |
|---|---|---|---|---|---|---|---|
| 1 | Female | 88 | 122.1 | 21.5 | | 124.2 | 1.72 |
| 1 | Female | 88 | 122.1 | 21.5 | | 127.3 | 4.26 |
| 1 | Female | 88 | 122.1 | 21.5 | | 141.0 | 15.48 |
| 1 | Female | 88 | 122.1 | 21.5 | | 142.7 | 16.87 |
| 1 | Female | 88 | 122.1 | 21.5 | | 146.4 | 19.90 |
| 1 | Female | 88 | 122.1 | 21.5 | | 149.2 | 22.19 |
| 1 | Female | 93 | 124.2 | 23.1 | | 127.3 | 2.50 |
| 1 | Female | 93 | 124.2 | 23.1 | | 141.0 | 13.53 |
| 1 | Female | 93 | 124.2 | 23.1 | | 142.7 | 14.90 |
| 1 | Female | 93 | 124.2 | 23.1 | | 146.4 | 17.87 |
| 1 | Female | 93 | 124.2 | 23.1 | | 149.2 | 20.13 |
| 1 | Female | 136 | 142.7 | 32.9 | | 146.4 | 2.59 |
| 1 | Female | 136 | 142.7 | 32.9 | | 149.2 | 4.56 |
| 1 | Female | 142 | 146.4 | 34.1 | | 149.2 | 1.91 |

Note: The figure illustrates an example of a prediction dataset, showing the inclusion of the target variable (Growth_Rate) alongside predictor variables. The age in months at the second observation (Months_T) is incorporated as a predictor variable for the first observation. This process is applied to all entities with two or more observations, creating a prediction dataset.

---

[1] The age variable was utilized in the model as months and labeled as 'Months'; however, for the convenience of readers it was converted to years and displayed as 'age (years)'.

Additional predictor variables were incorporated to effectively capture entities' growth dynamics and trajectories. Benchmark height (Height_BM) represents the projected height at Months_T, assuming the current height SDS remains constant. The corresponding growth rate at this point is defined as Growth_Rate_BM. Growth velocity variables were also incorporated, including height growth velocity (Height_V) and body composition growth velocities (Weight_V, Protein_V, SLM_V, BFM_V, SMM_V). These variables were calculated using data from observations obtained over the preceding 3-6 months, dividing the difference between current and past measurements by the time period in months.

The dataset with growth velocities comprised 105,774 measurements (54,463 for males and 51,311 for females) from 26,735 children and adolescents (13,907 males and 12,828 females). The final dataset input into the model was constructed by merging the target variable (Growth_Rate) and the prediction target information (Months_T, Growth_Rate_BM, and Height_BM) for each entity, resulting in 641,901 rows. The final dataset comprised 37 columns, consisting of the initial 27 attributes, 6 growth velocity metrics, 3 variables containing prediction target information, and the target variable itself. Of these 37 columns, 35 served as predictor variables. The sex variable and the target variable were excluded from predictors, with sex solely used for model differentiation.

Finally, distinct variable transformations were applied to the target and predictor variables separately. The target variable, height growth rate, was normalized using a power transformer with the Yeo-Johnson method to address positive skewness. A robust scaler was used for predictor variables to minimize the impact of outliers. Notably, due to these transformations, changes in the scaled target and predictor variables do not directly correspond to changes in their original units. The data preparation process is summarized in the following steps:

| Step | Details |
|---|---|
| **1. Initial Data Collection** | 276,301 measurements (54,374 subjects) |
| · Preprocessing original measurements | 145,292 males, 133,009 females |
|  | 27 attributes per measurement |
| **2. Prediction Dataset Creation** |  |
| · Identify multiple observations per subject |  |
| · Select two observations per subject (baseline and target) | 206,112 measurements (44,683 subjects) |
| · Calculate target variables (Growth_Rate) | 106,461 males, 99,651 females |
| · Construct a dataset with target variables |  |
| **3. Feature Engineering** |  |
| · Incorporation of benchmark features (Height_BM, Growth_Rate_BM) | 105,774 measurements (26,735 subjects) |
| · Computation of growth velocity variables | 54,463 males, 51,311 females |
| **4. Data Integration and Transformation** |  |
| · Merge target variables and prediction information | 641,901 rows |
| · Normalize target variable (Yeo-Johnson method) | 37 columns (35 predictors) |
| · Scale predictor variables (Robust scaler) |  |

*2.4. Prediction Model*

The height prediction model was trained using the light gradient boosting method (LightGBM), a gradient-boosting framework that employs tree-based learning algorithms. This framework is renowned for its efficiency and accuracy in processing large-scale datasets. The dataset division utilized a stratified sampling approach based on individual identification numbers. A random selection of twenty percent of individuals constituted the test set, while the remaining eighty percent formed the training set. This methodology ensured the exclusion of data from the same individual in both training and test sets, effectively preventing data leakage and mitigating potential overestimation of model accuracy. Separate models were developed for males and females to capture potential sex-specific differences in growth patterns. The representativeness and comparability of the training and test datasets were assessed by comparing the distributions of key variables, calculating the median and interquartile range (IQR) for each variable.

Hyperparameter optimization was performed using grid search with 5-fold cross-validation on the training set. The training set was further divided into 5 equal-sized folds, and the model was trained on 4 folds while validated on the remaining fold. This process was repeated 5 times, with each fold serving once as the validation set. The tuned hyperparameters included the maximum depth of the trees, the maximum number of leaves in each tree, the number of boosting rounds, and the learning rate. The best hyperparameter configuration from cross-validation was applied to train the final model on the entire training set. The trained model was then evaluated on an independent test set.

We evaluated the trained performances using the test set and computed three widely used performance metrics: root mean squared error (RMSE), mean absolute percentage error (MAPE), and mean absolute error (MAE). The model's stability and reliability were analyzed using a bootstrapping procedure, randomly splitting the test set into 50 subsets. Each subgroup's performance metrics (RMSE, MAE, and MAPE) were calculated, and the standard deviation (SD) and 95% confidence intervals (CI) were estimated for each metric. The heatmaps depicting MAPE values across various age groups and prediction horizons were generated to investigate the model's performance. Figure 2 depicts the model development pipeline flow chart from data preparation to model evaluation.

**Figure 2.** Model Development Pipeline Overview

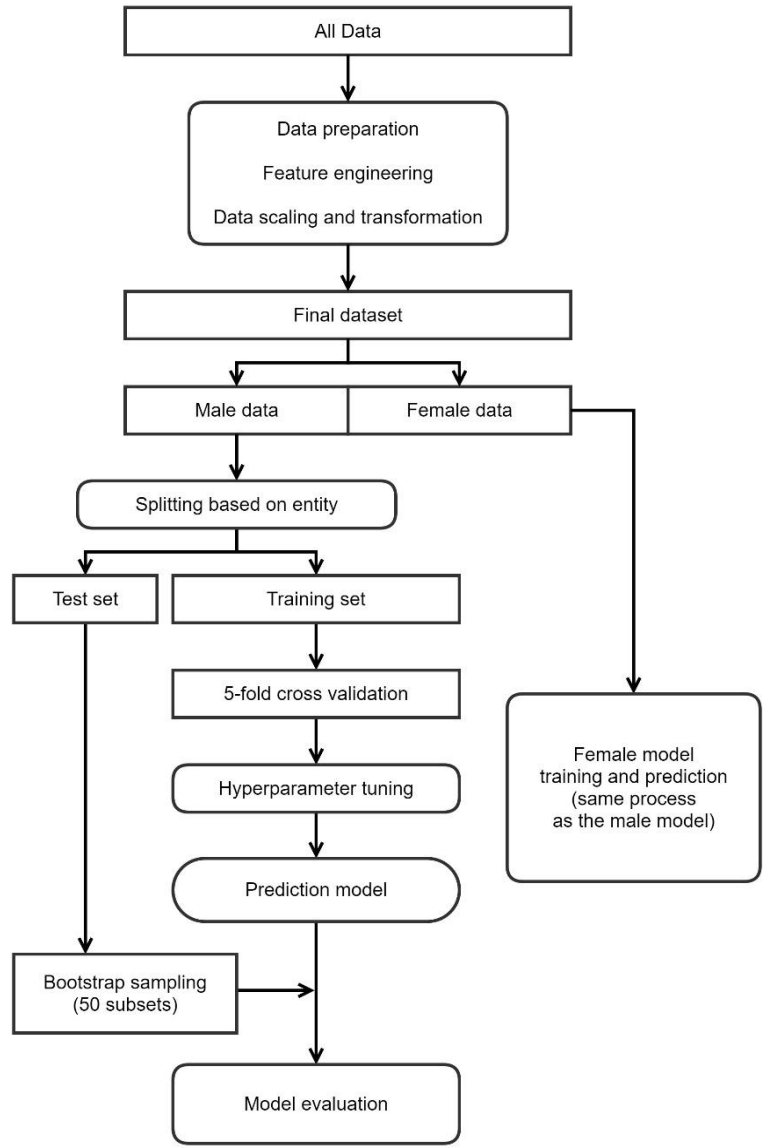

*2.5. Explainable AI*

The height prediction model was interpreted using three explainable AI techniques: feature importance, accumulated local effects (ALE), and SHAP. These methods illuminated key factors influencing predicted child and adolescent heights. Feature importance quantifies each feature's relative contribution to the model's predictions. The importance values were calculated based on the feature's average gain, with higher values indicating more influential features. For easier interpretation and comparison, these values were normalized to a scale of 100. ALE plots illustrate the average impact of individual features on model predictions across their value ranges, considering between-factor interactions. These plots demonstrate the cumulative local effect on the predicted height growth as a specific feature's value changes while other features remain constant. The analysis concentrated on prominent features determined by their importance scores. SHAP served as an additional method to interpret feature impacts on model predictions. Based on game-theoretic Shapley values, this approach calculates the marginal contribution of each feature to every prediction, enabling both global importance analysis and examination of individual cases. SHAP values were computed for each instance and visualized using summary plots.

*2.6. Personalized Growth Curve Estimation*

Personalized growth curve estimation forms a crucial component of our methodology, enabling individualized growth assessment. This approach generates a personalized growth trajectory by predicting heights at specific age points. The predicted curves are presented alongside actual height measurements and population-level references. Local SHAP was applied to confer model interpretability by identifying and highlighting the most influential features for individual predictions. To demonstrate these concepts, illustrative case studies of randomly selected male and female subjects are presented in the results section.

**3. Results**

*3.1. Baseline Characteristics*

Table 1 presents the study population's baseline characteristics, stratified by sex and dataset (training vs. test). For male subjects, the training set included 261,567 data points representing 11,125 unique individuals, while the test set contained 66,840 data points from 2,782 unique males. For females, the training set comprised 247,580 data points from 10,262 unique individuals, and the test set included 65,104 data points representing 2,566 unique females. The table presents the median and IQR for basic variables such as age (in years), height, weight, BMI, protein mass, SLM, BFM, SMM, BMR, TWB, and the target height.

In the training set, males had a median age of 9.08 years, height 134.9 cm, weight 32.3 kg, BMI 17.6 kg/m², protein mass 5.0 kg, SLM 23.9 kg, BFM 6.3 kg, SMM 13.0 kg, BMR 917 kcal, TBW 18.6 kg, and target height 147.8 cm. For females in the training set, the median age was 9.00 years, with a height of 133.6 cm, weight 30.6 kg, BMI 17.0 kg/m², protein mass 4.6 kg, SLM 22.2 kg, BFM 7.0 kg, SMM 12.0 kg, BMR 879 kcal, TBW 17.3 kg, and a target height of 148.3 cm. The test set showed similar values for both sexes, indicating an adequate balance between the training and test sets.

**Table 1.** Baseline Characteristics

|  | Male | | Female | |
| --- | --- | --- | --- | --- |
|  | Training set (n = 261,567) | Test set (n = 66,840) | Training set (n = 247,580) | Test set (n = 65,104) |
| Number of Entities | 11,125 | 2,782 | 10,262 | 2,566 |
| Age (months) | 9.08 [8.08-10.50] | 9.00 [8.17-10.42] | 9.00 [8.08-10.33] | 8.92 [8.08-10.33] |
| Height | 134.9 [128.6–142.8] | 135.0 [128.9–142.8] | 133.6 [127.3–142.5] | 133.2 [127.2–142.0] |
| Weight | 32.3 [27.2–40.3] | 32.3 [27.1–39.9] | 30.6 [25.9–37.5] | 29.8 [25.5–36.6] |
| BMI | 17.6 [16.0–20.2] | 17.5 [15.9–20.0] | 17.0 [15.5–19.1] | 16.7 [15.3–18.7] |
| Protein | 5.0 [4.4–5.8] | 5.0 [4.4–5.8] | 4.6 [4.1–5.5] | 4.6 [4.1–5.4] |
| SLM | 23.9 [21.1–27.9] | 23.9 [21.2–27.9] | 22.2 [19.7–26.2] | 21.9 [19.5–25.7] |
| BFM | 6.3 [4.0–11.0] | 6.0 [4.0–10.0] | 7.0 [5.0–10.0] | 6.0 [4.0–10.0] |
| SMM | 13.0 [11.3–15.6] | 13.0 [11.3–15.5] | 12.0 [10.4–14.4] | 11.8 [10.3–14.2] |
| BMR | 917 [853–1009] | 917 [855–1009] | 879 [821–971] | 873.0 [817–960] |
| TBW | 18.6 [16.4–21.7] | 18.6 [16.5–21.7] | 17.3 [15.3–20.4] | 17.1 [15.2–20.0] |
| Target Height | 147.8 [138.8–159.5] | 147.9 [138.8–158.9] | 148.3 [138.3–156.3] | 148.0 [138.1–156.0] |

The table presents the median and interquartile range [1st quartile–3rd quartile] for variables in the training and test sets. The number of data points (n) and unique individuals (Number of Entities) are also reported for each set.

*3.2. Prediction Accuracy*

Table 2 presents the prediction accuracy metrics, demonstrating high accuracy for both male and female models. For males, the models achieved mean RMSE of 2.51 ± 0.07 [95% CI 2.37 to 2.66] cm, MAE of 1.74 ± 0.05 [95% CI 1.65 to 1.83] cm, and MAPE of 1.14 ± 0.03 [95% CI 1.08 to 1.20]%. For females, the results showed mean RMSE of 2.28 ± 0.07 [95% CI 2.15 to 2.41] cm, MAE of 1.68 ± 0.05 [95% CI 1.58 to 1.78] cm, and MAPE of 1.13 ± 0.03 [95% CI 1.06 to 1.19]%.

**Table 2.** Overall Prediction Accuracy

| Sex | RMSE | MAE | MAPE |
| --- | --- | --- | --- |
| Male | 2.51 ± 0.07 | 1.74 ± 0.05 | 1.14 ± 0.03 |
|  | (2.37 to 2.66) | (1.65 to 1.83) | (1.08 to 1.20) |
| Female | 2.28 ± 0.07 | 1.68 ± 0.05 | 1.13 ± 0.03 |
|  | (2.15 to 2.41) | (1.58 to 1.78) | (1.06 to 1.19) |

The table presents the performance metrics for the height prediction models, including RMSE, MAE, and MAPE. Data are presented as the mean ± SD (95% CI) for each metric, allowing for an assessment of the model's stability and reliability across 50 different test data subsets.

To investigate the prediction model's performance, we generated heatmaps visualizing the MAPE across different age groups and prediction horizons for males and females (Figure 3). The heatmaps revealed that prediction accuracy varied depending on the current age and prediction horizon length. Generally, the model showed lower MAPE with shorter prediction horizons and older ages for both sexes. MAPE values increased with longer prediction horizons and younger ages, as evidenced by lighter colors in the upper-left corner of the heatmaps. Despite this variation, the model maintained robust performance, with the highest MAPE remaining below 2.42% for males (predicting 5 years ahead from age 7) and 2.00% for females (predicting 5.5 years ahead from age 10).

**Figure 3.** MAPE Heatmaps Across Different Age Groups and Prediction Horizons.

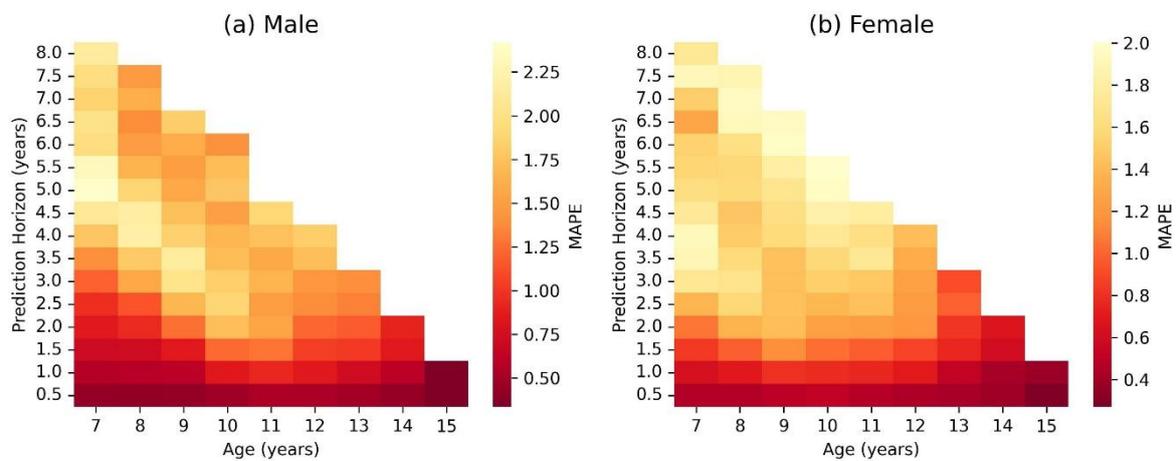

Notes: Heatmaps depicting MAPE across different age groups and prediction horizons for (a) males and (b) females. MAPE is calculated for each combination of age (in years) on the x-axis and prediction horizon (in years) on the y-axis. The color scale indicates the MAPE values, with darker shades of red representing lower errors.

*3.3. Feature Importance*

We identified the most significant height predictors by calculating feature importance values, which represent each feature's relative contribution to the prediction model based on its average gain (Figure 4). The top 20 features for each sex were visualized using horizontal bar plots, with importance values normalized to a scale of 100. We note that prediction target information such as Height_BM, Months_T, Months, and Growth_Rate_BM was excluded from this plot to focus on the anthropometric and body composition variables. The results showed that the SDS of current height (SDS_Height) was the most influential predictor for both males and females, followed by height velocity (Height_V) and soft lean mass velocity (SLM_V), after accounting for the prediction target information. Beyond these variables, additional top predictors for males included height, waist-to-hip ratio (WHR), and skeletal muscle mass (SMM). For females, other key predictors were waist-to-hip ratio, weight velocity (Weight_V), and height.

**Figure 4.** Feature Importance Plots.

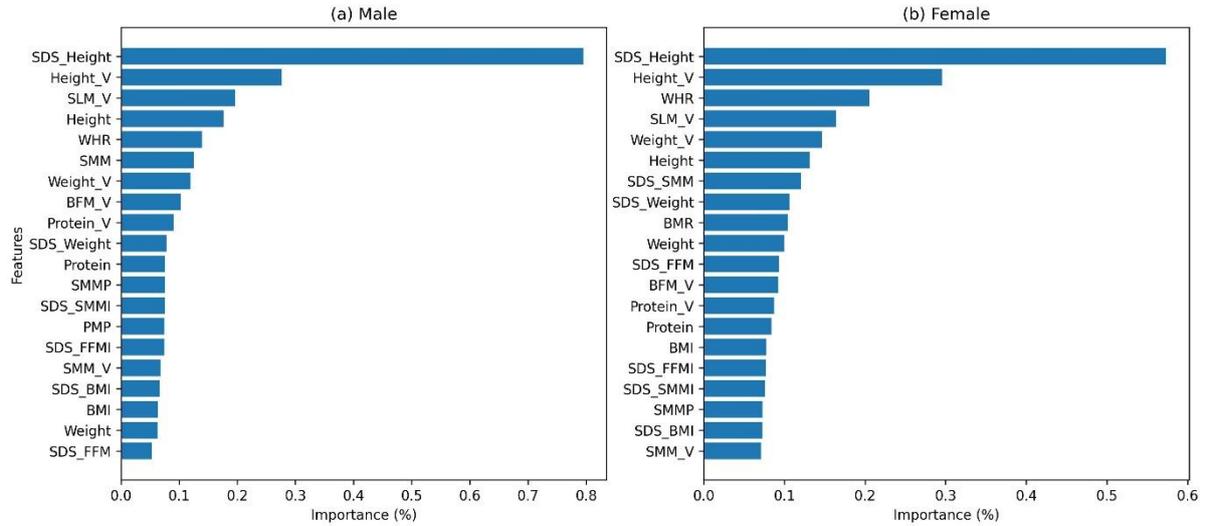

Notes: Importance of features within the height prediction models for (a) males and (b) females. The feature importance values are normalized to a total sum of 100, with higher values indicating greater importance. To focus on the importance of the anthropometric and body composition variables, we excluded prediction target information variables (Height_BM, Months_T, Months, and Growth_Rate_BM) from plots. Among these excluded variables, Growth_Rate_BM had an overwhelmingly high importance of 96.33% for males and 96.11% for females.

*3.4. ALE*

ALE plots visualize the average impact of individual features on model predictions while accounting for feature interactions (Figure 5). We find a robust negative and non-linear relationship between the SDS of current height (SDS_Height) and the predicted height growth. A clear negative correlation was evident when the scaled SDS_Height was less than 0, while no significant correlation was observed for values greater than 0. Similarly, a negative relationship was observed between height velocity (Height_V) and the predicted height growth. In contrast, soft lean mass velocity (SLM_V) is positively correlated with the expected height growth.

**Figure 5.** ALE Plots.

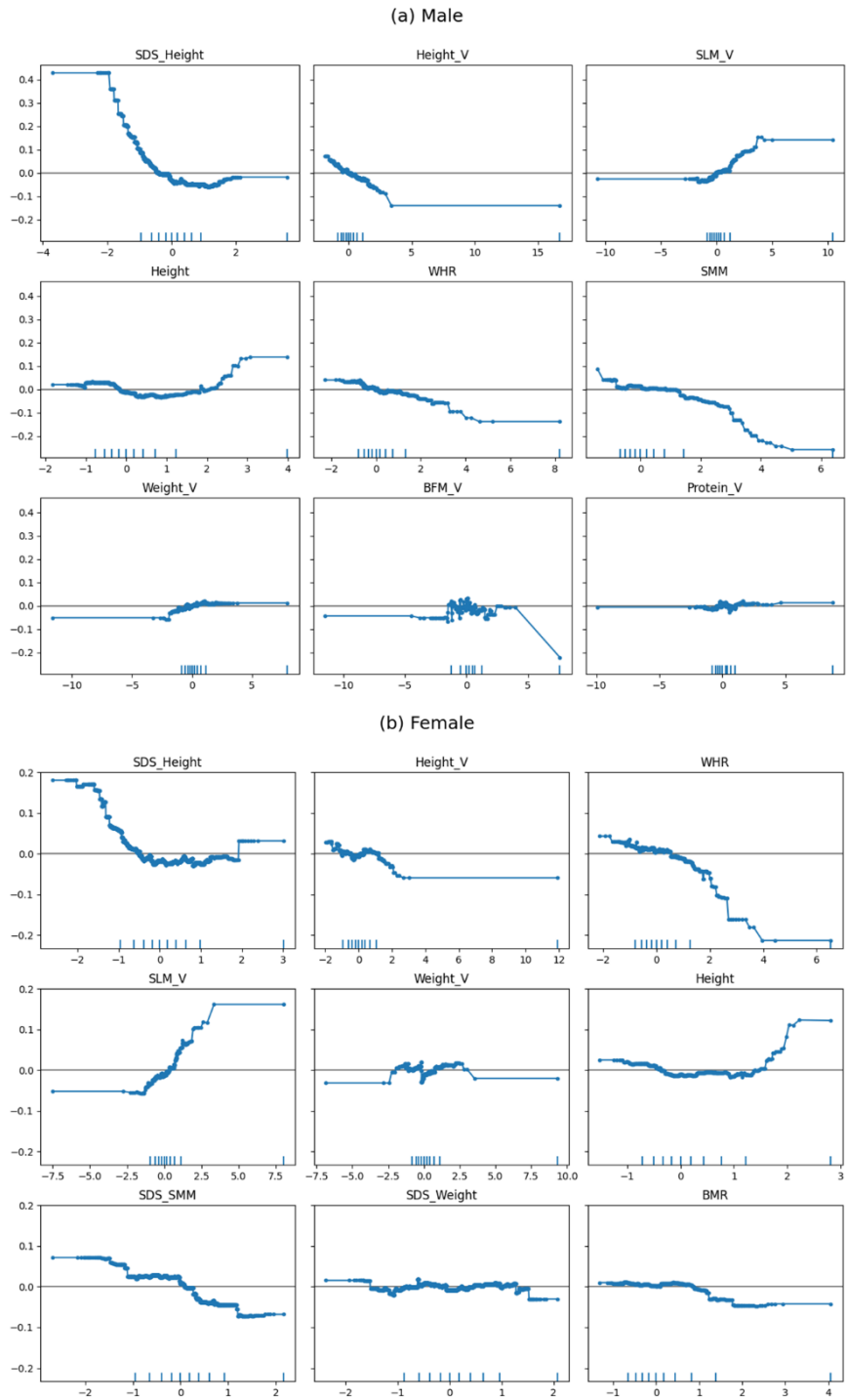

Notes: ALE plots for influential features in height prediction models for (a) males and (b) females, as determined by feature importance analysis. The figure illustrates the cumulative local effect on the predicted height growth as each feature value changes, while holding other features constant. The x-axis displays the range of values for the selected feature, while the y-axis indicates the corresponding change in the predicted height growth relative to the average prediction. Note that due to variable transformations, changes in the scaled target and predictor variables do not directly translate to changes in their original units.

*3.5. SHAP*

SHAP summary plots illustrate feature impacts on model predictions by quantifying Shapley values (Figure 6). Red indicates features pushing predictions higher, while blue indicates those pushing predictions lower. Key variables in both sexes include basic anthropometric measures like benchmark growth rate (Growth_Rate_BM) and current height SDS (SDS_Height). Growth velocities, such as soft lean mass velocity (SLM_V) and height velocity (Height_V), significantly influenced model predictions. In particular, height SDS and velocity showed inverse relationships with predicted growth rate, while soft lean mass velocity demonstrated a positive correlation, aligning with prior findings.

**Figure 6.** SHAP Summary Plots.

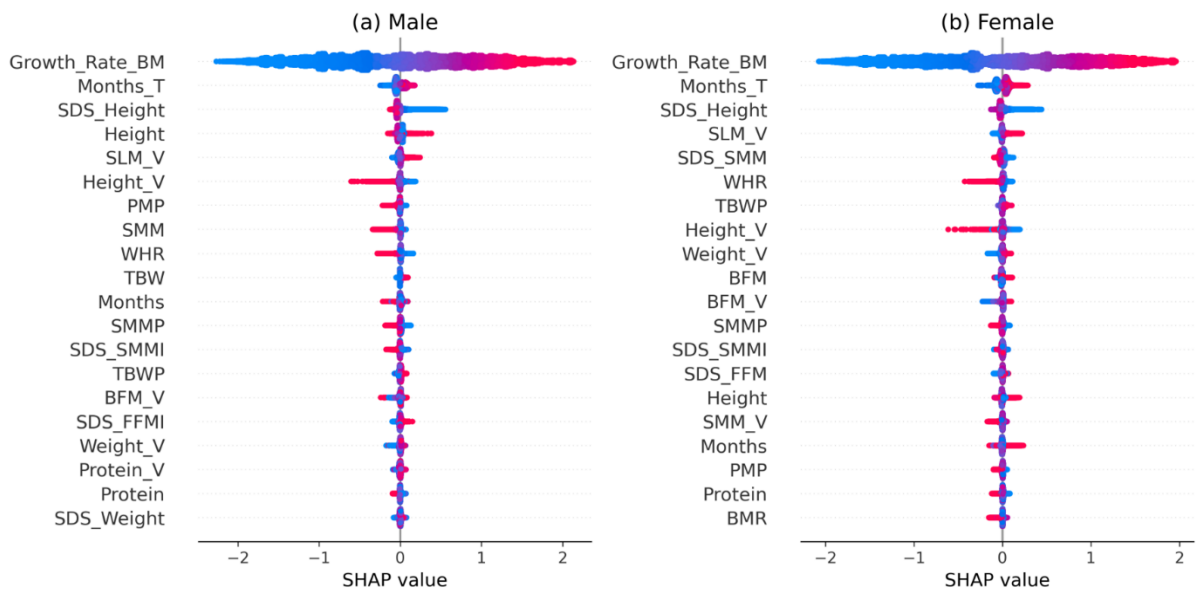

Notes: SHAP summary plots of models for predicting height in (a) males and (b) females. The plots illustrate the overall impact of each feature on the model's predictions, with features increasing the prediction shown in red and those decreasing the prediction shown in blue. Features are listed on the y-axis in descending order of importance. The x-axis represents the SHAP values, indicating the magnitude and direction of each feature's impact. The variable 'Months' represents the current age in months, while 'Months_T' indicates the target age in months.

*3.6. Individual Height Prediction Curves*

One of the key strengths of our height prediction model is its ability to generate personalized growth curves for individual children and adolescents. To demonstrate this capability and provide local interpretability, we present two illustrative examples (Figure 7). These examples were randomly selected from the test set, with one boy and one girl arbitrarily named Boy #1 and Girl #1, respectively.

**Figure 7**. Personalized Growth and SHAP Decision Plots.

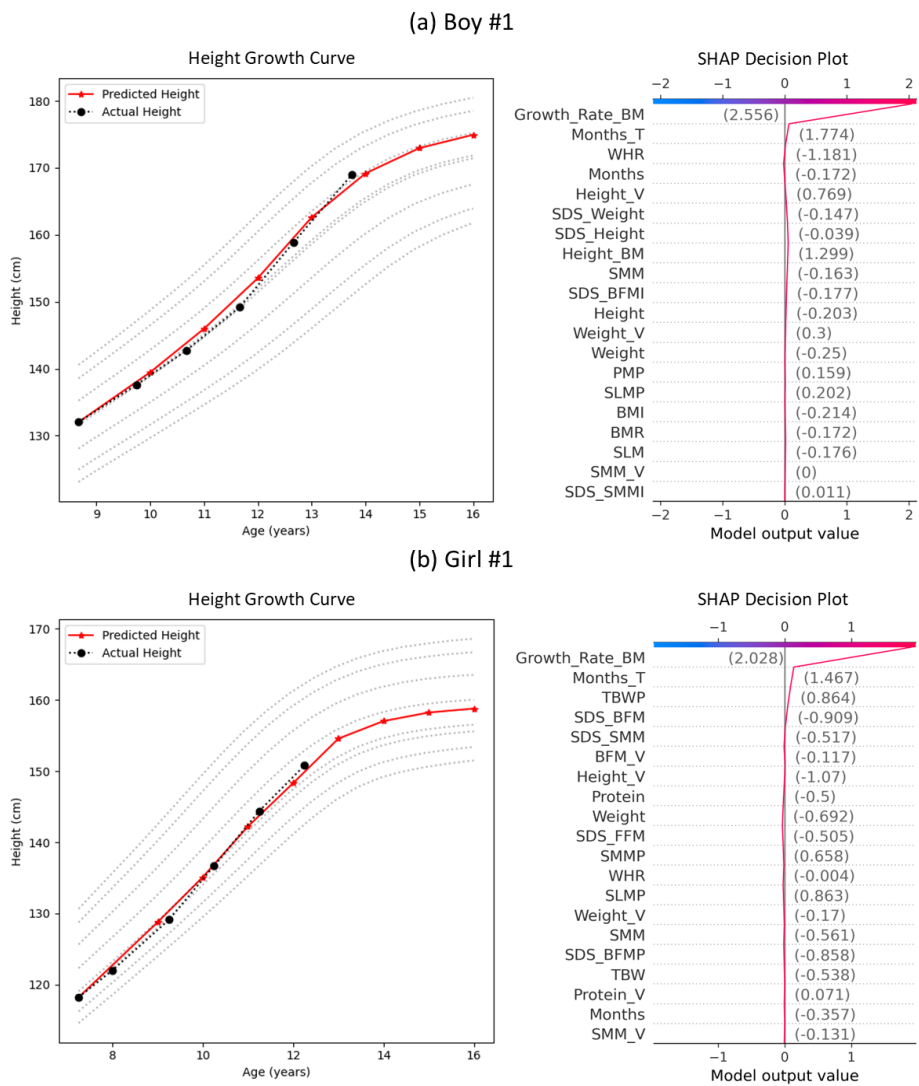

Notes: Personalized growth curves and SHAP decision plots for randomly selected (a) boy and (b) girl from the test set. The customized growth curves were generated using each child's longitudinal data. The SHAP decision plots quantify each feature's impact on the estimated height growth, with positive values indicating an increase and negative values a decrease in the predicted growth rate. The variable 'Months' represents the current age in months, while 'Months_T' indicates the target age in months.

Figure 7 displays predicted and actual height growth curves alongside corresponding SHAP decision plots for Boy #1 and Girl #1. These individual growth curves were generated using the trained model and available longitudinal data for each child. For Boy #1, the Benchmark growth rate (Growth_Rate_BM; 2.556) was the most significant positive contributor to the predicted height growth. Waist-to-hip ratio (WHR; 1.181) and height velocity (Height_V; 0.769) positively influenced the predicted height growth. In the case of Girl #1, the benchmark growth rate (Growth_Rate_BM; 2.028) was the most influential positive contributor. Furthermore, total body water percentage (TBWP; 0.864), SDS of body fat mass (SDS_BFM; −0.909), and SDS of skeletal muscle mass (SDS_SMM; −0.517) demonstrate positive influences on the predicted height growth.

**4. Discussion**

In this study, we developed an accurate and reliable AI model for predicting future height in children and adolescents, using comprehensive anthropometric and body composition data from a large-scale longitudinal study. The prediction model demonstrated mean RMSEs of 2.51 cm (males) and 2.28 cm (females), MAEs of 1.74 cm (males) and 1.68 cm (females), and MAPEs of 1.14% (males) and 1.13% (females). Performance varied across age groups and prediction horizons, with improved accuracy for shorter prediction periods and older age groups. Notably, the model maintained high accuracy across all scenarios, with MAPE values not exceeding 2.42% for males and 2.00% for females in any age group or prediction horizon.

Our results highlight the significant contributions of body composition and growth velocity to future height prediction. In particular, feature importance analysis and SHAP results identified height and SLM velocity as the most influential variables. The ALE plots revealed intriguing relationships between these growth velocities and the predicted height growth. A negative relationship between height velocity and predicted height growth aligns with the concept of catch-up growth, where children with prior growth delays may experience accelerated growth in subsequent periods (Boersma and Wit, 1997; Wit and Boersma, 2002; Albertsson-Wikland and Karlberg, 1994). In contrast, SLM velocity showed a positive correlation with predicted height growth, indicating that children with more muscle accretion tended to gain more height. However, the relationship between the increase in lean mass and future height in children and adolescents remains debatable. While these findings emphasize the complex link between lean mass growth and height attainment, the study's observational nature limits our conclusions to potential correlations rather than causal relationships.

The generation of personalized growth curves is a notable feature of this model. The integration of explainable AI techniques, particularly SHAP decision plots, provides interpretable insights into the factors driving individual growth predictions. For example, the SHAP decision plot for Boy #1 shows that his height prediction is largely influenced by his SDS of height, waist-to-hip ratio, and height velocity. In contrast, Girl #1's predicted height growth is primarily affected by her SDS of height, total body water percentage, and SDS of body fat mass. These personalized insights demonstrate the model's potential for tailored growth assessments in clinical settings.

Our study enhances pediatric growth assessment by applying ML algorithms to a comprehensive set of anthropometric and body composition measures. Conventional pediatric height predictions usually rely on skeletal maturity assessment using hand-wrist radiographs (Bayley & Pinneau, 1952; Tanner et al., 1975; Roche et al., 1975), which have limitations such as the need for specialized expertise, high interobserver variability, and reduced accuracy in assessing children with growth disorders (Bull et al., 1999; Chávez-Vázquez et al., 2024; Prokop-Piotrkowska et al., 2021; Topor et al., 2010; Bramswig et al., 1990; Bonfig & Schwarz, 2012). Our study provides complementary information by utilizing body composition data from BIA, which is non-invasive, cost-effective, and convenient, enabling large-scale and long-term data collection (Jansen et al., 2018; Hanem et al., 2019; Rodrígue-Carmona et al., 2022; Atazadegan et al., 2023). This method offers additional insights into biological maturity beyond skeletal maturity (Johnson et al., 2012; Dalskov et al., 2013; Jamison et al., 2006; Alimujiang et al., 2018; He and Karlberg, 2001; Brener et al., 2017).

To effectively utilize longitudinal body composition information, our study utilized SDS for body composition measures and growth velocities. Derived from a comprehensive database, these SDS offer a standardized approach to quantify relative growth status within age- and sex-matched cohorts (Chun et al., 2024b). This standardization enables the identification of growth patterns across diverse populations and enhances the comparability of growth data in longitudinal studies. Furthermore, our longitudinal data enables the calculation of growth velocities, which are crucial for identifying deviations from normal growth patterns (WHO 2009; Voss et al., 1991; McCarthy et al., 2007; Ekelund et al., 2006). The inclusion of body composition velocity provides unique insights into children's dynamic physical development, enhancing our understanding of growth trajectories.

Methodologically, this study applies ML approaches to large-scale biometric data for directly predicting height in children and adolescents. This framework addresses limitations in previous research, where ML applications in growth studies primarily focused on automated skeletal maturity assessment using radiographic data (Chávez-Vázquez et al., 2024; Martin et al., 2016; Thodbert et al., 2009, 2012; Suh et al., 2023; Satoh et al., 2015). Earlier attempts at direct height prediction were constrained by small dataset sizes and lack of comprehensive body composition measures (Mlakar et al., 2023; Rativa et al., 2018; Shmoish et al., 2021). This study addresses previous limitations by utilizing comprehensive body composition measures from the large-scale longitudinal GP Cohort Study. Employing specialized big data techniques, the model extracts information from a wide range of variables. This approach facilitates the development of an unbiased, objective, and reliable AI model that delivers consistent results.

The interpretable nature of our model enhances its potential clinical applications, offering valuable insights for both population-level analysis and individual patient care. At the population level, researchers and policymakers can identify risk factors and trends associated with suboptimal growth by analyzing key features driving growth predictions across large cohorts. For individual patient care, our model generates personalized growth curves, providing unique insights into growth patterns and developmental potential. Healthcare professionals can visualize predicted growth curves alongside actual measurements, assessing a child's growth trajectory in comparison to population-level reference curves. By combining personalized growth curve estimation with local interpretability, our model enables early identification of children at risk for growth disorders or deviations from expected trajectories, facilitating timely intervention and treatment monitoring (Collett-Solberg et al., 2019; Cuttler & Silvers, 2004; Topor et al., 2010; Bramswig et al., 1990; Bonfig & Schwarz, 2012). This transparency empowers clinicians to make informed decisions and communicate effectively with non-expert patients and their families.

A main limitation of our study is the lack of hand-wrist radiography data. This prevents us from directly comparing our ML-based approach with traditional methods for assessing skeletal maturity. If we had access to this data, we could have evaluated the additional value of our model to traditional methods and developed a more comprehensive framework for growth prediction. Another limitation is the study's observational nature; we could not strictly control factors that might influence growth outcomes. Consequently, the associations discovered between body composition and future height using explainable AI techniques should be interpreted carefully as they may not necessarily indicate direct causal relationships. Additionally, while the GP Cohort Study is a valuable resource for investigating growth patterns in South Korean children and adolescents, it is non-representative of other populations and ethnicities, including people living in different parts of the world. Moreover, our study focused on children and adolescents aged 7–16 years due to data availability, which may limit the applicability of our model to younger children or for predicting adult height. The accuracy of our model in these populations may be affected by the lack of data on early growth patterns, growth beyond age 16, and the potential influence of various factors on adult height. Furthermore, while our model demonstrates high accuracy and interpretability, integrating ML models into clinical workflows is challenging. To facilitate adoption, user-friendly interfaces and seamless integration with electronic health record systems will be essential. Healthcare professionals will also require training and support to effectively interpret and apply the model's insights to their clinical decision-making processes. Lastly, although our model incorporates a comprehensive set of anthropometric and body composition measures, it fails to account for other genetic, environmental, or socioeconomic factors that might affect growth.

Despite these limitations, our study offers valuable insights into the potential use of ML in pediatric growth assessment. By analyzing a large-scale longitudinal dataset and considering a wide range of factors, our model generates personalized growth curves and provides interpretable insights, which can be valuable for clinical decision making. Future research should prioritize collecting comprehensive data, including hand-wrist radiography, and employ rigorous experimental designs to validate and extend our findings. Our study marks a meaningful starting point for further exploration of using ML in pediatric growth assessment, emphasizing the importance of ongoing interdisciplinary collaboration to refine these techniques and implement them in clinical settings.

## 5. Conclusion

This study developed an ML model for predicting future height in children and adolescents, utilizing anthropometric and body composition measures. The model, trained on a large-scale longitudinal dataset, accurately predicts height and generates personalized growth curves. Explainable AI techniques were applied to enhance the model's interpretability. While the approach has limitations, it contributes to the field of pediatric growth assessment and may serve as a supportive tool in clinical settings, potentially aiding in growth disorder identification and outcome optimization.